\def\rn{\noindent\parshape 2 0truecm 8.8truecm 0.3truecm 8.5truecm}
\def\nn#1 #2{#1, #2.}				% Name with 1 initial
\def\nnn#1 #2 #3{#1, #2. #3.}			% Name with 2 initials
\def\nnnn#1 #2 #3 #4{#1, #2. #3. #4.}		% Name with 3 initials
\def\dualand{, \&\hbox{ }}				% Lower case "and" already in use.
\def\multiand{, \&\hbox{ }}				% Lower case "and" already in use.
\def\rfprep#1;#2;#3 {{\par\rn#1 #2, preprint #3\par}}
\def\rf#1;#2;#3;#4;#5 {\par\rn#1 #2, {\it #3}, {\bf #4}, #5\par}
\def\rfbook#1;#2;#3;#4;#5 {{\frenchspacing\par\rn#1 #2, {\it #3} (#4: #5)\par}}
\def\rfproc#1;#2;#3;#4;#5;#6 {{\frenchspacing\par\rn#1 #2, in {\it #3}, ed. #4 (#5: #6)\par}}

\def\etal{{\frenchspacing\it et al.}}
\def\ie{{\frenchspacing\it i.e.}}
\def\eg{{\frenchspacing\it e.g.}}

\def\cf{{\frenchspacing cf.}}

\def\beq#1{\begin{equation}\label{#1}}
\def\eeq{\end{equation}}
\def\beqa#1{\begin{eqnarray}\label{#1}}
\def\eeqa{\end{eqnarray}}
\def\eq#1{equation~(\ref{#1})}

\def\fig#1{Figure~1}
\def\Fig#1{Figure~1}

\def\sec#1{Section~\ref{#1}}

\def\spose#1{\hbox to 0pt{#1\hss}}
\def\simlt{\mathrel{\spose{\lower 3pt\hbox{$\mathchar"218$}}
     \raise 2.0pt\hbox{$\mathchar"13C$}}}
\def\simgt{\mathrel{\spose{\lower 3pt\hbox{$\mathchar"218$}}
     \raise 2.0pt\hbox{$\mathchar"13E$}}}
\def\simpropto{\mathrel{\spose{\lower 3pt\hbox{$\mathchar"218$}}
     \raise 2.0pt\hbox{$\propto$}}}

\def\Ms{M_{\odot}}     

\def\K{{\rm K}}
\def\meter{{\rm m}}
\def\second{{\rm s}}

\def\km{{\rm km}}

\def\keV{{\rm keV}}
\def\Ry{{\rm Ry}}
\def\Mpc{{\rm Mpc}}
\def\kpc{{\rm kpc}}

\def\tento#1{\times 10^{#1}}
\def\aet#1#2{\approx #1 \tento{#2}}

\def\zvd{z_{vd}}

\def\tvirfudge{f_{vir}}
\def\rhofudge{f_{\rho}}
\def\nstarfudge{f_\star}
\def\tvir{t_{vir}}
\def\nstar{n_{\star}}

\def\ed{\end{document}}

\documentstyle[emulateapj,epsf,rotate]{article}
\begin{document}

%%%%%%%%%%%%%%%%%%%%%%%%%%%%%

%\tighten
%\eqsecnum
%\received{4 August 1988}
%\accepted{23 September 1988}
\journalid{337}{15 January 1989}
\articleid{11}{14}

%\submitted{\today. To be submitted to ApJ.}
\submitted{ApJ, in press. Submitted September 8, 1997, accepted December 8.}

\title{WHY IS THE CMB FLUCTUATION LEVEL $10^{-5}$?}

\author{
Max Tegmark\footnote{Institute for Advanced Study, Princeton, 
NJ 08540; max@ias.edu}$^,$\footnote{Hubble 
Fellow}  
and
Martin J. Rees\footnote{Institute of Astronomy, University of Cambridge, 
Cambridge CB3 OHA, UK; mjr@ast.cam.ac.uk}$^{,1}$
}

\begin{abstract}

We explore the qualitative changes that would occur if the 
amplitude $Q\sim 10^{-5}$ of cosmological density fluctuations 
were different.
If $Q\simlt 10^{-6}$, the cosmological objects that form 
would have so low virial
temperatures that they may be unable to cool and form stars, 
and would be so loosely bound that even if they could produce a 
supernova explosion, they might be unable to 
retain the heavy elements necessary for planetary life.
If $Q\simgt 10^{-4}$, dense supermassive galaxies
would form, and biological evolution could be marred by 
short disruption timescales for planetary orbits. 
If $Q$ were still larger, most bound systems would collapse directly to
supermassive black holes.
These constraints on $Q$ can be expressed in terms of
fundamental constants alone, and depend only on
the electromagnetic and gravitational coupling constants,
the electron-proton mass ratio and the matter-to-photon ratio.
We discuss the implications for inflation and defect models, and
note that the recent anthropic upper bounds on the cosmological 
constant $\Lambda$ would be invalid if both $Q$ and $\Lambda$ 
could vary and there were no anthropic constraints on $Q$.
The same applies to anthropic bounds on the curvature parameter 
$\Omega$.
\end{abstract}

%%%%%%%%%%%%%%%%%%%%%%%%%%%%%%%%%%%%%%%%%%%%%%

\section{INTRODUCTION}

%  Because I want an "*", not a "4" in the "available" footnote:
\def\thefootnote{\fnsymbol{footnote}}
\footnotetext[1]{
{\it Available in color from}
{\bf h t t p://www.sns.ias.edu/$\tilde{~}$max/Q.html}
}
\def\thefootnote{\arabic{footnote}}

A key parameter in the standard adiabatic cold-dark matter-based models
of structure formation
is the amplitude that fluctuations in the gravitational potential
have when they enter the horizon. This number, which we will denote $Q$,
has been measured by the 
COBE satellite to be of order $10^{-5}$ 
(Smoot {\etal} 1992; Bennett {\etal} 1996), and is assumed to be virtually
independent of scale in the most popular models.
Why $10^{-5}$? The answers proposed by theorists fall into two categories:
\begin{enumerate}
\item $Q\sim 10^{-5}$ can be computed from first principles using some
(still undiscovered) fundamental theory.
\item $Q\sim 10^{-5}$ cannot be computed from first principles, since 
the correct fundamental theory merely predicts an ensemble of superhorizon-sized
spatial regions with a wide range of $Q$, forcing us to treat 
$Q$ as random number subject to various anthropic selection effects.
\end{enumerate}
%The purpose of this paper is to consider anthropic constraints 
%on the amplitude $Q$ of cosmological density fluctuations, which the COBE
%satellite has measured to be of order $10^{-5}$ 
%(Smoot {\etal} 1992; Bennett {\etal} 1996).
%Our motivation for this is threefold:
The purpose of this paper is to consider such selection effects, 
by studying how the physical processes of structure formation depend on $Q$.
Our motivation for this is threefold:
\begin{enumerate}
\item It affects which inflation/defect models should be considered natural
as opposed to fine tuned.

\item It is related to a crucial loophole in the recent arguments for
an anthropic upper bound on $\Lambda$.

\item It poses useful test problems for comparing cosmological simulations.

\end{enumerate}
The structure of our Universe is fixed by a rather small number of physical 
parameters. The electron mass, the neutron mass and the low energy coupling 
constants of the four basic forces determine the physical properties of
most objects on scales ranging from the atomic to the galactic 
(see {\eg} Carr \& Rees 1979; Davies 1982;
Barrow \& Tipler 1986), and these parameters can in 
turn be computed from the roughly 20
free parameters of the standard model of particle physics.
A number of additional parameters are often thought of as initial data
laid down in the early Universe: the baryon-to-photon ratio $\eta$, 
the relative abundances of various dark-matter candidates, the vacuum density
$\rho_\Lambda$ contributed by a cosmological constant $\Lambda$, the spatial 
curvature (related to $\Omega$) and the amplitude $Q$ of cosmological density 
fluctuations, although it is 
not implausible that abundances such as $\eta$ can ultimately be 
derived from other particle physics constants.
Together with the basic laws, these parameters determine when cosmic 
structures first emerge and how they evolve. Although the detailed 
outcome in any one locality, and what complex systems evolve there, 
depends on local accidents, these parameters nonetheless determine 
the statistical properties.

Will it ever be possible to compute the values of all these parameters from
first principles, within the framework of 
some yet to be discovered fundamental theory? 
The answer is a resounding no within some 
% plausible (albeit still speculative)
variants of inflationary cosmology 
(\eg, Linde 1983, 1987, 1990, 1995; Linde \& Zelnikov 1988; Coleman 1988; Albrecht 1994; 
Vilenkin 1995abcd; Vilenkin \& Winitzki 1997), 
where the spatial region that we conventionally call ``our Universe'', 
itself perhaps extending far beyond the present observational horizon, 
is just one element in an ensemble
whose members have widely disparate properties. 
Some physical parameters may take a range of different values
throughout this ensemble of exponentially large and causally 
disconnected regions.
The predictions of such theories therefore take the form of probability 
distributions for the parameters in question, and these must be
computed in Bayesian fashion taking into account the selection effect that
observers are not equally likely to inhabit all parts of the ensemble.
For instance, just as we expect low surface brightness galaxies 
to be underrepresented in many surveys, we might expect O-stars to be 
underrepresented in solar systems containing  
planet-based extraterrestrial civilizations and, 
as we shall see, spacetime regions with $Q\sim 10^{-20}$ to be
underrepresented in the set of regions that contain observers.
The importance of such anthropic selection effects was stressed 
by Carter (1974), and is discussed in great detail in books by, {\eg}, 
Davies (1982) and Barrow \& Tipler (1986).
More recent reviews can be found in, 
{\eg}, Balashov (1991) and Tegmark (1997).

% \subsection{Motivation 1: Inflationary predictions}
\subsection{Inflationary predictions}

Many inflationary models predict an ensemble of exponentially
large space-time regions, each with a different value of $Q$
(see {\eg} Linde 1990; Vilenkin 1995abcd and references therein).
Although the cosmological literature abounds with remarks on the 
``unnaturally'' flat potential required to produce $Q\sim 10^{-5}$
in our own Hubble volume, 
often as a motivation to study defect models, 
one can just as well argue that it is unnatural that the potential  
is not even flatter, since superflat potentials make inflation
last longer and hence dominate the ensemble by volume
(Vilenkin 1995a). This dispute cannot be resolved without taking
the inevitable anthropic selection effects into account: 
if these turn out to place a firm {\it upper} limit on 
$Q$ near the observed value, then inflation models
predicting ensembles peaked at {\it high} $Q$ clearly 
require no fine tuning to explain why we observe $Q\sim 10^{-5}$.
Conversely, if these selection effects give a firm lower limit on 
$Q$ near $10^{-5}$, then inflation models
predicting ensembles peaked at low $Q$ 
require no fine tuning.

%\subsection{Motivation  2: The cosmological constant puzzle}
\subsection{The cosmological constant puzzle}

Another hotly debated parameter is $\Lambda$, the cosmological constant.
Although one might expect the most ``natural'' value of the vacuum density 
$\rho_\Lambda$ to be of order the Planck density, the observational upper
limits on $|\rho_\Lambda|$ are a striking factor of $10^{123}$ smaller.
This has led to fine tuning criticism of cosmological models with
$\Lambda\ne 0$, the argument being that they were ruled out at high confidence, 
since such a small value of $\Lambda$ was extremely unlikely
(see Dolgov 1997 for an up-to-date review).
As was pointed out by Barrow \& Tipler (1986), 
Weinberg (1987, 1989) and Efstathiou (1995), 
there is a flaw in this argument, 
since it neglects a powerful anthropic selection effect. 
If $\Lambda$ is too large, then the Universe becomes vacuum 
dominated before the density
fluctuations have grown enough to form non-linear structures. Hence the fluctuations 
stop growing, and neither galaxies nor observers will ever form.
It is therefore no surprise that we find ourselves in a region where 
$\Lambda$ is small. A calculation of the probability distribution for 
$\rho_\Lambda$ given our existence shows that values of order of
the current limits are in fact rather typical (Efstathiou 1995), and 
more accurate calculations (Weinberg 1996; Martel {\etal} 1997)
have confirmed this conclusion.

Unfortunately, there is a loophole in this argument (Rees 1997). As described in
more detail in Section 5, increasing $\Lambda$ by some  
factor $f$ can be completely offset by increasing $Q$
by a factor $f^{1/3}$ as far as this argument is concerned.
Whether this is really a loophole thus depends crucially on 
the topic of the present paper, specifically on whether  
observers could exist if $Q\gg 10^{-5}$.
The analogous potential loophole exists for anthropic lower bounds on
$\Omega$ ({\cf} Barrow 1982; Vilenkin \& Winitzki 1997).

%\subsection{Motivation 3: Simulation-testing}
\subsection{Simulation-testing}

A third and entirely different motivation for exploring
counterfactual values of parameters such as $Q$ is that it
provides a challenging and bias-free test of cosmological 
simulation techniques. State-of the art simulations 
including hydrodynamics (which breaks the degeneracy
between $Q$ and $t$ in pure gravity simulations), 
gas chemistry and star formation
often achieve a good fit to our actual universe
(see {\eg} Kang {\etal} 1994 and references therein), but only after 
tweaking a number of parameters empirically. It is therefore unclear
to what extent the agreement between different groups is 
due to realistic modeling as opposed to simply living in
(and parameter-fitting to) the same Universe.
It would be far more convincing if two groups could obtain 
indistinguishable results for hypothetical universes with other 
values of $Q$, where the answer would not be known beforehand.

In Section 2, we
outline how $Q$ affects structure formation in a universe
with $\Omega=1$ and $\Lambda=0$.
We discuss the effects of lowering and raising $Q$ in Sections
3 and 4, respectively, and
the effects of changing $\Omega$ and $\Lambda$ in Section~5.

% Perhaps explain why we're simply using using PS instead of 
% the complicated stuff Martel did: it's hardly worth being more elaborate 
% than that, since the nonlinear calculation isn't all that realistic anyway.

\section{GALAXY FORMATION AND COOLING}

\subsection{Notation}

We will find it convenient to work in 
Planck units where $\hbar=c=G=k=1$, and the fundamental units of 
of length, time, mass and temperature are 
$r_{pl}\equiv(\hbar G/c^3)^{1/2}\aet{2}{-35}$m, 
$t_{pl}\equiv(\hbar G/c^5)^{1/2}\aet{5}{-44}$s,
$m_{pl}\equiv(\hbar c/G)^{1/2}\aet{2}{-8}$kg and
$T_{pl}\equiv(\hbar c^5/G)^{1/2}/k\aet{1}{32}$K,
respectively.
Important dimensionless constants that will recur frequently are
the electromagnetic coupling constant
$\alpha\equiv e^2\approx 1/137$, the 
gravitational coupling constant $\alpha_g\equiv m_p^2\aet{6}{-39}$,
the electron-proton mass-ratio $\beta\equiv m_e/m_p\approx 1/1836$,
the baryon-to-photon ratio $\eta\sim 10^{-9}$, 
the baryon fraction $\Omega_b/\Omega\sim 10^{-1}$
of the nonrelativistic matter 
density (which we take to equal the critical value that gives a spatially
flat Universe) and the matter-to-photon ratio
\beq{xiDefEq}
\xi\equiv 
{m_p\eta}{\Omega\over\Omega_b}=
{\alpha_g^{1/2}\eta}{\Omega\over\Omega_b}
\sim 10^{-27}.
\eeq
This constant 
$\xi$ is simply the amount of nonrelativistic
matter per photon, $\rho_m/n_\gamma$, 
measured in Planck masses.
As our goal is to highlight the main physical effects rather than 
to make detailed numerical calculations,
we will frequently use the symbol $\sim$, which we take to
mean that numerical factors of other unity ($\pi$ and the like)
have been omitted. For instance, the hydrogen binding energy 
(1 Rydberg), the Bohr radius and the Thomson cross section  
are given by $\Ry\sim\alpha^2\alpha_g^{1/2}\beta$, 
$a_0=\alpha^{-1}\alpha_g^{-1/2}\beta^{-1}$ and
$\sigma_t\sim\alpha^{2}\alpha_g^{-1}\beta^{-2}$, respectively.

The reader may find it unfamiliar to see almost no reference 
below to familiar quantities such as the redshift $z$, 
the current CMB temperature $T_0\approx 2.726\K$, the 
current Hubble constant $H_0$ and the current density 
parameter $\Omega_0$. This is because  
we strive to highlight how structure formation depends on 
fundamental parameters, and these quantities are not fundamental 
since they have meaning only once the epoch at which we happen to
be living has been specified.
Indeed, for the the open Universe case, $T_0$, $H_0$ and $\Omega_0$  
can be thought of as merely alternative time variables, 
since they all decrease monotonically with $t$. For instance, we are  
not interested in examining what $Q$-values allow galaxies to form 
by the present epoch $t_0\sim 10^{10}$ years, but 
what $Q$-values allow them to form at all.

\subsection{When non-linear structures form}
\label{FormationEpochSec}

The rising curves in \fig{tMfig} show when different mass scales 
go nonlinear, defined as the time when linear perturbation 
theory predicts an overdensity of 1.69 in a top hat sphere 
containing the mass $M$ (Press \& Schechter 1974).
The curves were computed for the 
cold dark matter (CDM) power
spectrum fit of Bond \& Efstathiou (1984) with $h=0.5$,
``shape parameter'' $\Gamma=0.25$, and
an $8h^{-1}\Mpc$ normalization $\sigma_8=0.7\times (Q/10^{-5})$.
We assume a standard spatially flat Universe
($\Omega=1$, $\Lambda=0$) everywhere in this paper\footnote{
We assume a standard scale-invariant Harrison-Zel'dovich primordial
power spectrum throughout this paper.
More general primordial spectra would correspond to a scale-dependent $Q$,
thus requiring more than a single number for their parametrization.}
except in 
\sec{OmegaLambdaSec}.
Since fluctuations cannot grow before the matter-radiation equality 
epoch\footnote{
At $t_{eq}$, the radiation energy
per proton, $T_{eq}/\eta$, equals the dark matter energy 
per proton, $m_p\Omega/\Omega_b$, so $T_{eq}\sim m_p\eta\Omega/\Omega_b=\xi$.
Since the energy density is $\rho_{eq}\sim T_{eq}^4$, 
the Friedman equation gives the Hubble expansion rate
$H\sim\rho^{1/2}\sim T_{eq}^2$, and so the age of Universe at this time
is $t_{eq}\sim H^{-1}\sim T_{eq}^2\sim\xi^{-2}$.
%is $\t_{eq}\sim H^{-1}\sim T_{eq}^{-2}\sim\alpha_g^{-1}\xi^{-2}$.
}
$t_{eq}\sim\xi^{-2}\sim 10^{11}\,$s 
(the vertical line in the figure), all scales 
below the horizon mass at this epoch,
\beq{MeqEq}
M_{eq}\sim \rho_{eq} t_{eq}^3\sim \xi^{-2}
\sim \alpha_g\xi^{-2}\Ms \sim 10^{16}\Ms,
\eeq
have similar fluctuation levels, and are seen to virialize roughly simultaneously
(up to a logarithmic factor), at 
\beq{tvirEq}
\tvir\sim t_{eq} Q^{-3/2}\tvirfudge \sim\xi^{-2}Q^{-3/2}\tvirfudge. 
\eeq
(The origin of the ``$3/2$'' is that, during the matter-dominated epoch, 
fluctuations grow as the scale factor $a$ and $a\propto t^{2/3}$.)
Since the figure shows that the actual curves approach vertical 
only for very small mass scales, we have included a factor $\tvirfudge$ 
in \eq{tvirEq} which depends weakly on mass. 
$\tvirfudge\sim 1$ for $M\sim M_{eq}$, with the value for typical
galactic scales $M\sim 10^{12}\Ms$ being $\tvirfudge\sim 0.03$. % zzzz fusk
Far above this mass scale, $P(k)\simpropto k$ 
(we assume the standard spectral index $n=1$), which means that 
%\beq{MvirEq}
$M\sim M_{hor} Q^{3/2}$,
%\eeq
where the horizon mass is $M_{hor}\equiv t$    % c^3t/G$ 
(straight solid line). 
Thus the curves all have the same shape, and their 
left and right asymptotes lie about a factor $Q^{-3/2}$ to the right of
the two heavy straight lines in the figure, 
giving the following broad-brush picture.
Mass scales $M\simlt M_{eq}$ virialize roughly simultaneously, 
at $t\sim t_{vir}$. As time progresses, ever larger scales
keep virializing, the non-linear mass scale always being a fraction 
$Q^{3/2}$ of the horizon mass scale (a fraction $Q^{1/2}$ in radius).
Note that the the number $10^{16}$ occurring in this crucial mass
$M_{eq}$ is simply $\alpha_g/\xi^2$ --- the well-known result that 
a stellar mass $\Ms\sim \alpha_g^{-1}$ 
(Dyson 1971)
%(Dyson 1971; Weisskopf 1975)
was used in \eq{MeqEq}.

\subsection{Their virial temperature}

When an overdensity has collapsed, the resulting virial halo 
will have a typical density that exceeds the background density
by a collapse factor $\rhofudge\sim 18\pi^2$, \ie,
\beq{rhovirEq}
\rho_{vir}\sim\rho_{eq}\left({t_{vir}\over t_{eq}}\right)^{-2}\rhofudge
\sim\xi^4 \tvirfudge^{-2}\rhofudge Q^3.
\eeq
For a CDM halo of mass $M$, this corresponds to a characteristic
size $R\sim (M/\rho_{vir})^{1/3}$,
velocity $v_{vir}\sim (MG/R)^{1/2}\sim (M^2\rho_{vir} G^3)^{1/6}$ 
and virial temperature 
\beq{TvirEq}
T_{vir}\sim m_p v_{vir}^2  % \sim M^{2/3}\rho_{vir}
\sim\alpha_g^{1/2} \xi^{4/3}  \tvirfudge^{-2/3}\rhofudge^{1/3} M^{2/3} Q.
\eeq
A number of isotherms are plotted in \fig{tMfig}, and we see that as
time progresses and ever larger halos form, the
virial temperature stops increasing around the characteristic time
$t\sim\tvir\propto Q^{-3/2}$ and approaches a maximum value
$T_{max}\sim m_p c^2 Q$, corresponding to a maximum virial velocity
$v\sim Q^{1/2} c$. Thus for our $Q\sim 10^{-5}$ universe, 
typical cluster temperatures
are $\sim 10\>\keV$, about $10^{-5}$ times the proton rest energy,
and characteristic cluster velocity dispersions are 
$1000\>\km/\second$, about $10^{-5/2}$ times the speed of light.

\section{WHAT IF $Q\ll 10^{-5}$?}

This direct link between $Q$ and halo temperatures 
immediately indicates why lowering $Q$ can cause
qualitatively different structure formation scenarios.
Unless $m_p c^2 Q$ exceeds typical atomic 
energy scales $\sim 1\>\Ry$, 
which corresponds to 
\beq{QionEq}
Q \simgt {\Ry\over kT_{max}}
\sim{ m_e c^2\alpha^2\over m_p c^2} =\alpha^2\beta\sim 10^{-8},
\eeq
it will be difficult for the
gas in these halos to dissipate their energy to
collapse and form stars. 
Hydrogen line cooling freezes out at about
$\Ry/15\sim 10^4\K$, for instance, 
corresponding to $Q\sim 10^{-9}$.
We will now discuss cooling constraints in more detail, and
see that these cause qualitative changes even for much smaller 
departures from $Q\sim 10^{-5}$.

\begin{figure}[phbt]
{{\vbox{\epsfxsize=16cm\epsfbox{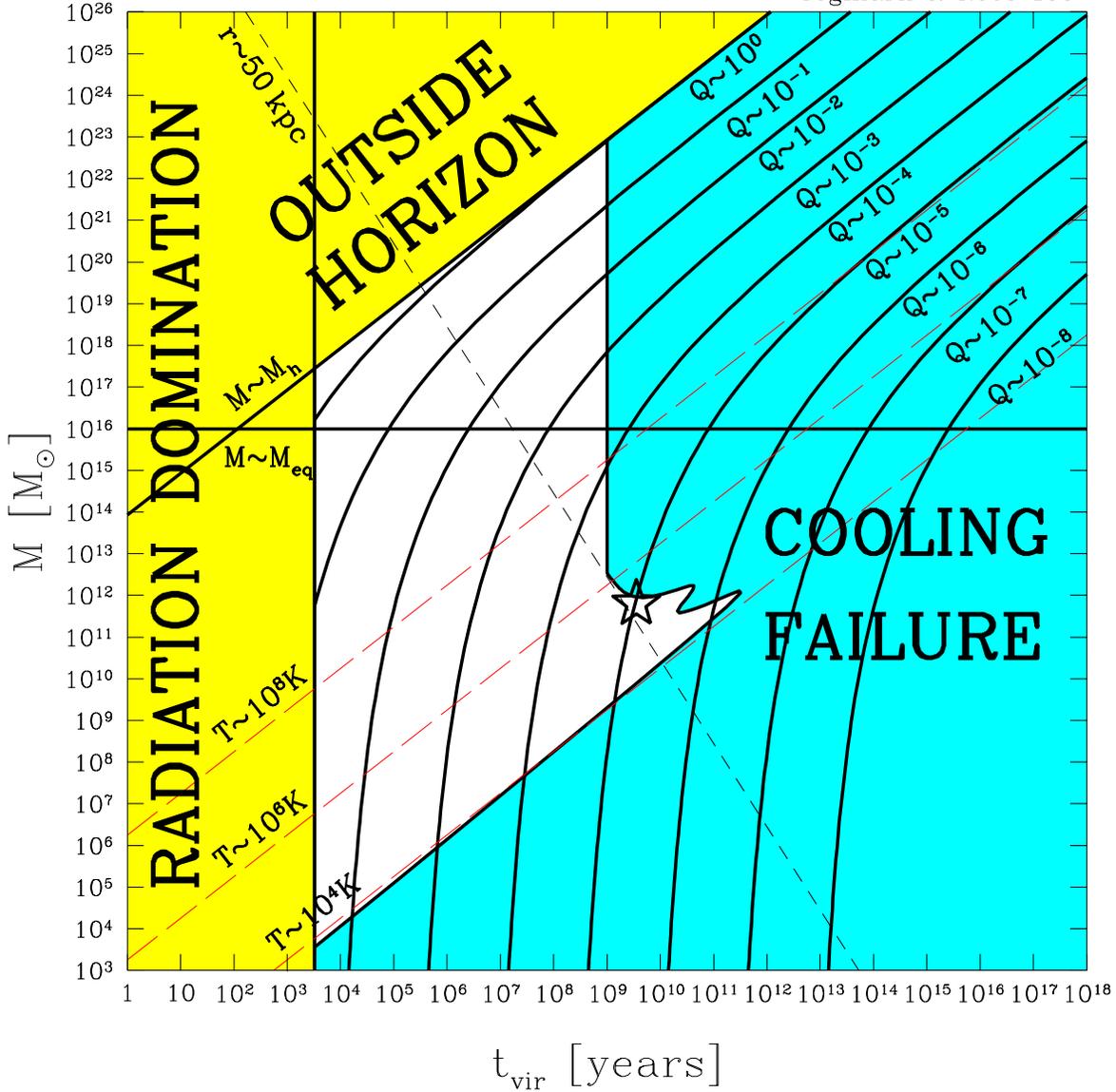}}}}
\caption{
%The effect of altering $Q$.
The nine rising curves show the largest virialized mass scale
as a function of time for different values of $Q$.
Structures with $M\simlt M_{eq}$ (horizontal line)
are seen to all virialize about a factor $Q^{-3/2}$ after 
the end of the radiation-dominated epoch (shaded, left), 
whereas for later times, the virialized
mass scale asymptotes to about $Q^{3/2}$ times
the horizon mass (shaded, upper left).
Cooling is inefficient in the remaining 
shaded region (right). The star corresponds to 
the Milky Way.
%The five isotemperature curves are 
%$10^n$ K, where $n=$2, 4, 6, 8 and 10.
%The downward-sloping line is for  your ``magic radius''
%$\sim$ 50 kpc.
%I've shaded regions that are ruled out by
%(1) being outside the horizon, (2) being before equality,
%(3) being unable to cool in a local Hubble time.
\label{tMfig} 
}
\end{figure}

The fate of the baryons in a virialized halo depends crucially
on the ratio of the cooling timescale $\tau_{cool}\equiv T/\dot T$
to the  gravitational collapse timescale 
$\tau_{grav}\sim (\rho_{vir} G)^{-1/2}$
(see {\eg} Binney 1977, Rees \& Ostriker 1977; 
Silk 1977; White and Rees 1978).
If $M$ and $t$ are such that $\tau_{cool}\simgt\tau_{grav}$ 
(the dark-shaded region in \fig{tMfig}), the cloud 
cannot promptly commence free-fall collapse and fragment into stars,
but will remain pressure supported for at least a 
local Hubble time.  
For the halo formation curve corresponding to $Q\sim 10^{-5}$,
the part of the $\tau_{cool}=\tau_{grav}$ curve setting the
upper limit on galaxy mass is seen to have a logarithmic
slope around $-2$ (because Bremsstrahlung,
with $\tau_{cool}\propto T^{1/2}/\rho$, is the
dominant cooling process), 
corresponding to $M\propto\rho\propto t^{-2}$ and a 
constant radius 
$R\sim\alpha^3\alpha_g^{-3/2}\beta^{-3/2}\sim 50\,\kpc$
(Carr \& Rees 1979). The corresponding mass
scale is seen to be $M\sim 10^{12}\Ms$. 
For slightly lower $Q$, the upper limit
is dominated by line cooling in neutral Hydrogen
(rightmost bump), Helium (second bump) and any heavier
elements released by early stars (not included here).
The lower mass limit is set by the $T\sim 10^4\K$
isotherm, below which there are essentially no free electrons
and both line cooling and Bremsstrahlung become ineffective.
Molecular cooling can potentially lower
this mass limit slightly ({\cf} Haiman {\etal} 1996;
Abel {\etal} 1997; Gnedin \& Ostriker 1997; Tegmark {\etal} 1997), 
but is ignored in the figure for the same reason as
heavy elements: it is irrelevant 
to our $Q$-constraints, which depend only how far the cooling 
region extents to the right, not on the vertical extent.

What happens if we start lowering $Q$? 
The first change is that the upper limit becomes set 
not by Bremsstrahlung but by line cooling.
\Fig{tMfig} indicates
that as we keep lowering $Q$, the range of galactic masses
narrows down and finally vanishes completely for
$Q<Q_{min}\sim 10^{-6}$. % zzzz fusk  
Let us express this critical value $Q_{min}$ in fundamental constants.
The figure shows that it is determined by the
``Hydrogen bump'' in the cooling function, 
which is caused by free electrons collisionally
exciting neutral Hydrogen atoms into their first 
excited state, which is immediately followed by
emission of a Ly$\alpha$ photon. 
This gives a cooling timescale ({\eg}, 
Dalgarno \& McCray 1972)
\beq{LineCoolEq}
t_{cool}\sim \left({m_e^2 c\alpha\over\hbar^2}\right)
{\gamma^{-3/2} e^{{3\over 4}\gamma}\over x(1-x)n},
\eeq
where $\gamma\equiv \Ry/kT$, $n$ is the total
(bound and free) proton number density, 
and $x$ is the ionization fraction.
In thermal equilibrium, this is given by
({\eg} Tegmark, Silk \& Evrard 1993)
\beq{xEq}
x\sim[1+\alpha^3\gamma^{7/6}e^\gamma]^{-1}.
\eeq
Substituting \eq{xEq} into \eq{LineCoolEq} gives
\beq{LineCoolEq2}
t_{cool}\sim
\left({m_e^2 c\over\hbar^2\alpha^2 n}\right)
\left[\gamma^{-8/3} e^{-\gamma/4}
(1+\alpha^3\gamma^{7/6}e^\gamma)^2\right],
\eeq
where the dimensionless 
quantity in square brackets is minimized
for $\gamma\sim\ln[\alpha^{-2}]\sim 10$, corresponding 
to $T\sim\Ry/10\sim 15,000\K$. This minimum value
is $\sim\gamma^{-8/3}e^{-\gamma/4}\sim
\alpha^{1/2}\ln[\alpha^{-2}]^{-8/3}\sim 1/5,000$.
Equating this minimal cooling timescale with 
$t\sim(G\rho)^{-1/2}$ using $n=\rho\Omega_b/m_p$ 
finally tells us that the latest time at which line cooling
can be efficient is 
\beq{tmaxEq}
t_{max} \sim \alpha^{3/2}\ln[\alpha^{-2}]^{8/3}
\alpha_g^{-3/2}\beta^{-2}\Omega_b\sim 10^{19}\second.
\eeq
Equating this with $t_{vir}$ from \eq{tvirEq} thus tells us that 
efficient cooling occurs when  
\beq{QlineEq}
Q\simgt \alpha^{-1}\ln[\alpha^{-2}]^{-16/9}
\alpha_g\beta^{4/3}\xi^{-4/3}\tvirfudge^{2/3}\Omega_b^{-2/3}
\sim 10^{-6}.
\eeq 

If $Q\ll 10^{-6}$, then what is the ultimate fate of
the quasistatic pressure supported gas clouds? 
It is plausible that they will become increasingly rarefied
as their dark matter halos eventually merge into larger 
(and less dense) halos, thereby never entering a phase
of runaway cooling, fragmentation and star formation.
However, even in the arguably contrived case where 
such a cloud
escaped any further collisions, and eventually 
managed to cool after a 
(perhaps exponentially) long time, perhaps 
through some exotic mechanism such as 21 cm cooling,
and developed a dense, self-gravitating core which 
fragmented into stars, there would still be reason to doubt
whether it could produce intelligent observers.
Since the binding energy of the halo is so low
(of order $T_{vir}$), the first supernova explosion
might well eject all the gas from the halo, 
thereby precluding the production 
of population II stars and planets containing heavy elements.
%In general, if star formation is a rare event occurring at
%intervals $\gg t_{max}$, greatly exceeding stellar lifetimes, 
%it may be very

\section{WHAT IF $Q\gg 10^{-5}$?}

What happens if we start increasing $Q$ instead? 
The allowed mass range for galaxies keeps broadening at a steady
rate until Compton cooling suddenly eliminates the upper mass limit 
altogether. This is because the time scale on which 
cosmic microwave background (CMB) photons at temperature 
$T_\gamma$ cool an ionized plasma,
\beq{taucompEq}
\tau_{comp}\sim {\hbar^3 c^4 m_e\over \sigma_t(k T_\gamma)^4 x},
\eeq
is independent of both its density and temperature 
(assuming that $T\simgt 15,000\>\K$, so that $x\sim 1$).
Since $T_\gamma\sim T_{eq}(t/t_{eq})^{-2/3}\sim\xi^{-1/3}t^{-2/3}$, 
this timescale $\tau_{comp}$ 
equals the age of the universe $t$ at a characteristic time
\beq{tcompEq}
t_{comp}\sim\alpha^{6/5}\alpha_g^{-13/10}\beta^{-9/5}\xi^{-4/5}\sim10^{16}\second.
\eeq
Setting $t_{comp}=t_{vir}$, we find that the upper limit
to galaxy masses persists only for
\beq{QcompEq}
Q\simlt\alpha^{-4/5}\alpha_g^{3/5}\beta^{6/5}\xi^{-4/5}\tvirfudge^{2/3}\sim 10^{-4.5}.
\eeq
For larger $Q$-values, all mass scales can cool efficiently, 
so the characteristic mass for the first generation of 
galaxies will simply be $M_{eq}\sim 10^{16}\Ms$, given by
\eq{MeqEq}. This corresponds to a characteristic size
$R\sim t_{eq}/Q\propto t_{eq}t^{2/3}$ for newly formed
galaxies,
which is constant in {\it comoving} coordinates
(rather than in absolute coordinates, as the above-mentioned 
cooling scale $R\sim 50\>\kpc$).
It would plainly need detailed simulations to determine the mix of discs and
spheroids, and the effects of subsequent mergers. However, the galaxies could
well have a broader luminosity function than in our actual universe (as well as
a much higher characteristic mass); and clustering would also extend up to a
larger fraction of the Hubble radius.

\subsection{Disruption of planetary orbits}

Would this qualitative change affect the number of habitable 
planets produced?
Let us first consider the stability of planetary orbits.\footnote{
After submission, it was brought to the authors' attention that Vilenkin 
discussed orbit disruption caused by high Q-values at the 1995
Tokyo RESCEU symposium in Tokyo. Orbit disruption constraints 
on galaxy densities have also been discussed in the context 
of axion physics (Linde 1998).}

Lightman (1984) has shown that if the planetary surface temperature is
to be compatible with organic life, the orbit around the central
star should be fairly circular and have a radius of order
\beq{rAUeq}
r_{au}\sim\alpha^{-5}\alpha_g^{-3/4}\beta^{-2}\sim 10^{11}\meter,
\eeq
roughly our terrestrial ``astronomical unit'',
precessing one radian in its orbit on a timescale
\beq{tYearEq}
t_{orb}\sim\alpha^{-15/2}\alpha_g^{-5/8}\beta^{-3}\sim 0.1 \hbox{ year}.
\eeq
An encounter with another star with impact parameter $r \simlt r_{au}$
has the potential to throw the planet into a highly eccentric orbit
or even unbind it from its parent star. This happens on a timescale
$\tau_{enc}\sim 1/\nstar v r_{au}^2$, where $\nstar$ and 
$v\sim v_{vir}$ 
denote the typical stellar density and stellar velocity in a galaxy,
respectively. 
Writing $\nstar\sim\nstarfudge\rho_{vir}/M_{\star}$, where 
$M_{\star}\sim\alpha_g^{-1}$ and $\nstarfudge$ is
the additional factor by which the dissipating baryons collapse
relative to the dark matter before fragmenting into stars,
the Milky Way is empirically fit by $\nstarfudge\sim 10^1$.
For Earth, this gives $\tau_{enc}\sim 10^{22}\second$, orders of
magnitude above its present age. 
Moreover, the distant encounters that we have experienced in the 
past have had a completely negligible effect since they were
adiabatic. This means that the impact duration 
$r/v\gg t_{orb}$, so that the solar system
returned to its unperturbed state once the encounter 
was over.
For hypothetical galaxies forming 
before $t_{comp}$, on the other hand, 
$M\sim M_{eq}$, so the time between non-adiabatic encounters is 
\beq{tAdiabEq}
\tau_{adiab}\sim{1\over\nstar v^3 t_{orb}^2}\sim
{\alpha^{15}\alpha_g^{1/4}\beta^{6}\tvirfudge^3\over\xi^4 Q^{9/2}\rhofudge^{3/2}\nstarfudge}
\sim 10^{5}\hbox{ years}
\eeq
for $Q=10^{-4}$ and 
dropping as $Q^{-9/2}$ if we increase $Q$ further. In other words,
non-adiabatic encounters are frequent events for $Q\simgt 10^{-4}$, 
occurring often during the geological timescales required for
a planet to form, cool and ultimately evolve life.
In the conservative approximation of ignoring gravitational
focusing (assuming that the flyby speed $v$ exceeds the orbital speed),
the typical time interval between $r<r_{au}$ encounters is 
\beq{tEncEq}
\tau_{enc}
\sim{1\over\nstar v r_{au}^2}
\sim
{\alpha^{10}\alpha_g^{1/2}\beta^{4}\tvirfudge^{7/3}\over\xi^{4}\rhofudge^{7/6}\nstarfudge Q^{7/2}}
\sim 10^{7}\hbox{ years}
\eeq
for $Q=10^{-3}$.
Requiring this to exceed some geological or evolutionary timescale 
$t_{min}$ thus gives an upper limit $Q\propto t_{min}^{-2/7}$.
Although it is far from clear what is an appropriate $t_{min}$
to use, the smallness of the exponent $2/7$ implies that it makes
only a minimal difference whether we choose $10^6$ or $10^{10}$ years.
Taking $t_{min}\sim 10^9$ years $\sim\alpha^2\alpha_g^{-3/2}\beta^{-2}$,
the lifetime of a bright star (Carr \& Rees 1979), we obtain the limit
\beq{QdissEq}
Q\simlt 
\alpha^{16/7}\alpha_g^{4/7}\beta^{12/7}\xi^{-8/7}\tvirfudge^{2/3}\rhofudge^{-1/3}\nstarfudge^{-2/7}
%\alpha^{10/9}\beta^{5/3}\alpha_g^{1/9}\xi^{-4/3}\Omega_b^{-2/9}\tvirfudge^{2/3}\rhofudge^{-1/3}
\sim 10^{-4}.
\eeq
This upper limit appears more uncertain than the lower limit from cooling.
The momentum kick given to the planet scales as $v^{-2}$, so an impact with
$r\ll r_{au}$ would not necessarily cause a catastrophic disturbance
of the planetary orbit --- the event rate for this grows only as $Q^{5/2}$,
or as $Q^3$ if the galactic stars settle into a disk where 
$v$ is roughly independent of $Q$. On the other hand, a very close 
encounter (especially with an O-star) might cause disastrous heating 
of the planet. In view of this uncertainty, as well as the uncertainty 
regarding $\nstarfudge$ and $f_{min}$, we now consider two additional
effects of raising $Q$.

\subsection{Black hole domination}

For much greater $Q$-values, of order
unity, typical fluctuations would be of black-hole magnitude already by the time
they entered the horizon, converting some substantial fraction $f$ of the
radiation energy into black holes already shortly after $t_{infl}$, 
the end of the inflationary era. At $t_{eq}$, the universe has expanded by
a factor $a\sim(t_{eq}/t_{infl})^{1/2}$, and the energy densities in black holes
and photons have dropped by factors of $a^3$ and $a^4$, respectively.
The black hole density will therefore completely dwarf the density 
of cold dark matter and baryons if $f\gg a^{-1}$. Thus even if $Q\ll 1$, 
extremely rare fluctuations that are $Q^{-1}$ standard deviations out in 
the Gaussian tail can cause black hole domination if 
$\Phi(Q^{-1})\sim a^{-1}$, where 
$\Phi(x)\equiv(2\pi)^{-1/2}\int_{x}^\infty\exp[-u^2/2]du$.
This gives the upper limit
\beq{QbhEq}
Q\equiv\Phi^{-1}[\xi]
\sim\Phi^{-1}[10^{-27}]\sim 10^{-1},
\eeq
where we have simply assumed that $t_{infl}$ is within a few orders of 
magnitude of unity (the Planck time) since this affects the result only
logarithmically. As opposed to the previous constraints, this one 
depends strongly on whether the power spectrum is strictly scale invariant 
or not --- increasing the spectral index $n$ from its scale-invariant value
$n=1$ to $n=1.3$ causes primordial black hole domination even if 
$Q$ is as low as $10^{-5}$ (Green {\etal} 1997).

%\subsection{Black hole formation during matter domination}

Even if $Q$ were low enough to avoid black hole formation in the early
radiation dominated phase (say in the range $10^{-3}$--$10^{-2}$), rampant
black-hole formation may still occur in the matter-dominated era.
At times of order $10^6$ years, {\ie}, shortly after recombination, 
clumps of order $M_{eq}$ will collapse. If dissipation leads to enough
reionization to make their Thomson optical depth larger 
than $c/v$ (itself of order $Q^{-1/2}$), then they will
trap the background radiation and collapse like supermassive stars, 
without being able to fragment. 
The dominant structures in such a universe 
would then be supermassive black holes,
and it is unclear whether any galaxies and stars would be able 
to form. Even if they could, they would be
hurtling around at speeds of order a tenth of the speed of light,
and it is far from clear how anthropically favorable such an 
environment would be!

%If $Q$ were still larger, of order $10^{-2}$, then the first giant galaxies
%would form before the end of the Compton drag epoch, which means that
%they would trap radiation as they collapsed and perhaps form supermassive
%black holes whose subsequent accretion could douse the remaining matter 
%in intense high-energy radiation. 

%* It seems that if you assume flatness, you by definition get 
%$\Lambda<1$, which is rather
%dull. So nice to include Omega too for this reason.

\section{What if $\Lambda$ and $\Omega$ were different?}
\label{OmegaLambdaSec}

Our discussion above applied to a flat FRW universe with
$\Omega=1$ and $\Lambda=0$. As we will now describe, anthropic
limits on these two parameters are intimately linked with $Q$.
In Planck units, the Friedmann equation that governs the time evolution of
the radius of curvature of the Universe, $a$, is conveniently 
written as 
\beq{FriedmannEq}
\left({\dot a\over a}\right)^2 = 
{8\pi\over 3}(\rho_\gamma+\rho_m+\rho_c+\rho_\Lambda),
\eeq
where $\rho_\gamma$, $\rho_m$ and $\rho_\Lambda$
are the energy densities corresponding to radiation, 
nonrelativistic matter and vacuum energy (a cosmological constant), 
respectively. $\rho_c\equiv\pm 3/8\pi a^2$ is the 
contribution from spatial curvature (the sign is 
positive if $\Omega<1$ and negative if $\Omega>1$ --- for 
the flat case $\Omega=1$, the radius of curvature is infinite
and $a$ must be redefined).
The first three of these densities evolve as
\beqa{densityEq}
\rho_\gamma&\sim&\rho_\Omega^{-2}a^{-4},\\
\rho_m&\sim&\xi\rho_\Omega^{-3/2}a^{-3},\\
\rho_c&\sim&a^{-2},
%\rho_\Lambda&=&\rho_\Lambda a^0.
\eeqa
and $\rho_\Lambda$ does not evolve at all.
The constant $\rho_\Omega$ is defined as the 
curvature that the Universe {\it would} have had at the Planck time
if there was no inflationary epoch, and can be evaluated at 
any time in the post-inflationary radiation-dominated epoch
as $\rho_{\Omega}=\rho_c t\sim t/a^2$, during which this quantity
is time independent. We have introduced $\rho_\Omega$ simply because we need
a constant that quantifies the curvature, and the more familiar 
$\Omega$ is unusable since it changes with time.
The epochs of matter domination $a_{md}$, 
curvature domination $a_{cd}$ and vacuum domination
$a_{vd}$ are given by $\rho_\gamma\sim\rho_m$, 
$\rho_c\sim\rho_m$ and
$\rho_\Lambda\sim\rho_m$, respectively, {\ie}, 
\beqa{EqualityEq}
a_{md}&\sim&\xi^{-1}\rho_\Omega^{-1/2},\\
a_{cd}&\sim&\xi\rho_\Omega^{-3/2},\\
a_{vd}&\sim&\xi^{1/3}\rho_\Omega^{-1/2}\rho_\Lambda^{-1/3}.
\eeqa
It is well-known that sub-horizon fluctuations can only grow 
during the matter-dominated epoch, where they grow at the same rate
as the scale factor $a$.
As we saw in \sec{FormationEpochSec}, the first non-linear structures
therefore form at $a_{vir}\sim a_{md}Q^{-1}$ providing that
the Universe remains matter-dominated until this epoch
($a_{cd}\simgt a_{vir}$ and $a_{vd}\simgt a_{vir}$)
--- otherwise no nonlinear structures will ever form.
We thus obtain the two anthropic constraints
\beqa{OmegaLimitEq}
\rho_\Omega&\simlt&\xi^2 Q\sim 10^{-59},\\
\label{LambdaLimitEq}
\rho_\Lambda&\simlt&\xi^4 Q^3\sim 10^{-124}.
\eeqa
Although we tacitly assumed that $\Omega<1$ here, the closed
case gives essentially the same constraints --- indeed, 
if no non-linear structures have formed at the epoch $a_{cd}$
in a closed universe, time is literally running out for 
not yet evolved life forms, since the Big Crunch is imminent!

In comparison, the current observational limits are
(very conservatively) $\rho_\Lambda\simlt\rho_m\sim 10^{-123}$ and 
$0.1\simlt\Omega\simlt 2$, which corresponds to 
$a_{vd}\simgt 10^3 a_{md}$ and 
$\rho_\Omega\simlt10^{-57}$.
The conclusion is that although the anthropic upper limits superficially
appear quite strong on both curvature and vacuum density, 
these constraints are only strong
if the two variables on the right-hand side 
($\xi$ and $Q$) are independently 
constrained --- which was one of our motivations for studying
upper limits on $Q$. 
%Extensive reviews of 
%limits on $\alpha_g$ can be found in, 
%{\eg}, Barrow \& Tipler (1986), Balashov (1991) and Tegmark (1997).
The parameter $\xi$ probably deserves more attention
than it has received in this context so far ({\eg} Rees 1979),
and the effects of varying the baryon/photon ratio 
and introducing a non-zero neutrino mass would also 
warrant further study.
We note in passing that we can obtain crude $Q$-independent limits 
on $\xi$ by requiring that our lower limits on $Q$ not exceed our
upper limits. For instance, the virialization epoch of \eq{tvirEq} will  
occur too late for cooling to be efficient (after $t_{max}$ of  
\eq{tmaxEq}) unless $\xi\simgt 10^{-32} Q^{-3/4}$.
Thus the white region in the figure disappears completely
if $\xi\simlt 10^{-32}$, and the 
conservative limit $Q\simlt 10^{-3}$
gives the (rather weak) constraint $\xi\simgt 10^{-30}$.
Conversely, the planetary disruption constraint of \eq{QdissEq} gets
stronger if we {\it increase} $\xi$, and conflicts with the
$\xi$-independent limit of \eq{QionEq} unless 
$\xi\simlt 10^{-23}$.
In addition, there are of course separate limits on the baryon fraction
$\Omega_b$, in that if there 
are too few baryons, the cooling becomes less efficient 
--- see \eq{tmaxEq}.
Lowering $\Omega_b$ may also impede
galaxy and star formation, since a gas cloud must collapse by a larger
factor before it becomes self-gravitating.

For the reader preferring to think in terms of $\Omega_0$ and  
redshift $z$, the above argument can be re-expressed as follows.
If the current matter density is $\rho_m$, then vacuum domination
occurs at the epoch $(1+\zvd)=(\rho_\Lambda/\rho_m)^{1/3}$.
If $\Omega_0\ll 1$, then the Universe became 
curvature dominated at a redshift given by $(1+z_{cd})\sim\Omega_0^{-1}$.
Since the first structures form at an epoch $(1+z_{vir})\propto Q$, 
the upper limits on $\Lambda$ and $\Omega_0^{-1}$  
thus scale as $\Lambda\propto Q^3$ and $\Omega_0\propto Q^{-1}$ 
for the $\Omega_0\ll 1$ case. 
% and the anthropic constraints in the 
% $(\Omega,\Lambda)$-plane are roughly degenerate along the
% curve $\Lambda\propto\Omega^{-3}$ for the $\Omega\ll 1$ case.
For instance, maintaining spatial flatness but 
making $\Lambda$ a million times larger than the current
observational limits could correspond
to $Q\sim (10^6)^{1/3}\times 10^{-5} = 10^{-3}$, 
with galaxy formation about ten expansion times after 
recombination. When the Universe had reached its current age of $\sim 10^{10}$ 
years, it would have expanded by a further
factor $\sim e^{100}$, and ours would be the 
only galaxy in the local Hubble volume --- alas, a drab and dreary place 
for extragalactic astronomers, but not ruled out by the above-mentioned
$\Lambda$-arguments alone --- although perhaps by the $Q$-arguments that we
have presented.

\section{DISCUSSION}

We have explored counterfactual cosmological scenarios with  
$Q$ shifted away from its observed value $\sim 10^{-5}$.
We found that qualitative changes occur if we either
increase or decrease $Q$ by about an order of magnitude.
If $Q\simlt 10^{-6}$, efficient cooling becomes impossible
for gas in virialized halos. If $Q\simgt 10^{-4}$, 
Compton scattering against CMB photons 
enables efficient cooling in arbitrarily
massive halos, and the higher stellar densities
and velocities may lead to planetary orbits being disrupted
before observers have had time to evolve.

Needless to say, this does not preclude that some form of
life might evolve in a Universe with a 
more extreme $Q$-value due to lucky
circumstances, for instance around a field star that was ejected 
from its giant host galaxy in a $Q\sim 10^{-3}$ scenario.
However, as stressed by Vilenkin (1995a), the key feature
of anthropic selection effects is not what the rock-solid 
extreme limits are on a parameter, but which is 
{\it the most favorable value} for producing observers.
This point is also emphasized by {\eg} Garc\'\i a-Bellido \& Linde (1995).
To predict a probability distribution for the observed value of $Q$
from some inflationary model (to potentially rule the model out),
its a priori probability distribution for $Q$ (of quantum origin, say)
must be multiplied by some Bayesian selection function such as 
the number of observers or civilizations corresponding to each $Q$-value.
It seems plausible that much more stars with habitable planets are formed 
for $Q\sim 10^{-5}$
(where perhaps $1\%-10\%$ of all baryons are in stars) than in 
a $Q\sim 10^{-6}$ universe where 1000 times lower densities
make cooling difficult. Likewise, it appears likely that 
$Q\sim 10^{-4}$ gives fewer planets in
favorable stable orbits than $Q\sim 10^{-5}$, where close encounters
are completely negligible for most stars.
In conclusion, it is possible that the anthropic selection
function peaks at $Q\sim 10^{-5}$. If this is the case, then 
what Vilenkin terms 
``the principle of mediocrity'' would imply that since we are 
most likely to be a
typical civilization, this is what we should expect to observe.

%So far, no inflation or defect model has been put forward that
%can make the unique prediction $Q\sim 10^{-5}$ from first principles.

The authors wish to thank Tom Abel, Andreas Albrecht,
Ted Bunn, Wayne Hu, Robert Kirshner, Andrei Linde, Avi Loeb, 
Scott Tremaine, Alexander Vilenkin and 
Ned Wright for helpful suggestions. % conversations
This work was supported by
NASA grant NAG5-6034 and Hubble Fellowship
{\#}HF-01084.01-96A, awarded by the Space Telescope Science
Institute, which is operated by AURA, Inc. under NASA
contract NAS5-26555.

%\clearpage

%\end{document}

%%%%%%%%%%%%%%%%%%%%%% FIGURES: %%%%%%%%%%%%%%%%%%%%%%%%%

\end{document}